\documentclass[12pt]{article}
\usepackage{pic04}
\usepackage{hyperref}
\usepackage{url}
\usepackage{graphicx}

\begin{document}

\title{\bf JET QUENCHING AT RHIC}
\author{
Saskia Mioduszewski        \\
{\em Physics Department,
Brookhaven National Laboratory,
Upton, NY 11973}}
\maketitle

%
%
%
%
%
%
\vspace{4.5cm}
%

\baselineskip=14.5pt
\begin{abstract}
We present high transverse momentum ($p_T$) measurements made at the
Relativistic Heavy Ion Collider (RHIC) for Au+Au, d+Au, and p+p collisions 
at $\sqrt{s_{NN}} = 200$~GeV, as well as for Au+Au collisions at
$\sqrt{s_{NN}} = 62$~GeV.
We observe a suppression in the yield of high $p_T$ hadrons measured
in central Au+Au collisions, relative to the yield in 
p+p collisions scaled by the number of binary nucleon-nucleon collisions.
This observation, together with the absence of such a suppression in
d+Au collisions, leads to the conclusion that a dense medium is formed
in central Au+Au collisions.
\end{abstract}

\newpage

\baselineskip=17pt

\section{Heavy Ion Collisions}

Numerical simulations of quantum chromodynamics (QCD) on a lattice 
predict a phase transition (or crossover) from hadronic matter 
to deconfined, chirally symmetric matter at sufficiently large energy 
densities~\cite{lattice}.  The critical energy density calculated is typically
$\sim 0.7$~GeV/fm$^3$, approximately 5 times the density of normal
nuclear matter.
The primary goal of high energy heavy ion physics is to achieve such a 
transition and to study this new state of matter, the Quark Gluon 
Plasma (QGP), in order to better understand fundamental properties
of QCD.  With the new collider RHIC, 
the field has recently reached a new energy regime, which will be
a milestone in the endeavor to attain this goal.

The Relativistic Heavy Ion Collider (RHIC) 
is located at Brookhaven National 
Laboratory (BNL) on Long Island, New York.  RHIC has provided Au+Au 
collisions at $\sqrt{s_{NN}} = 130$~GeV; Au+Au collisions 
at $\sqrt{s_{NN}} = 62$~GeV; and Au+Au, d+Au, and 
p+p collisions at $\sqrt{s_{NN}} = 200$~GeV.  At these energies, 
hard scattering is expected to 
contribute significantly to particle production.

\subsection{Hard Scattering in Heavy Ion Collisions}

\begin{figure}[hbt]
\includegraphics[width=9cm]{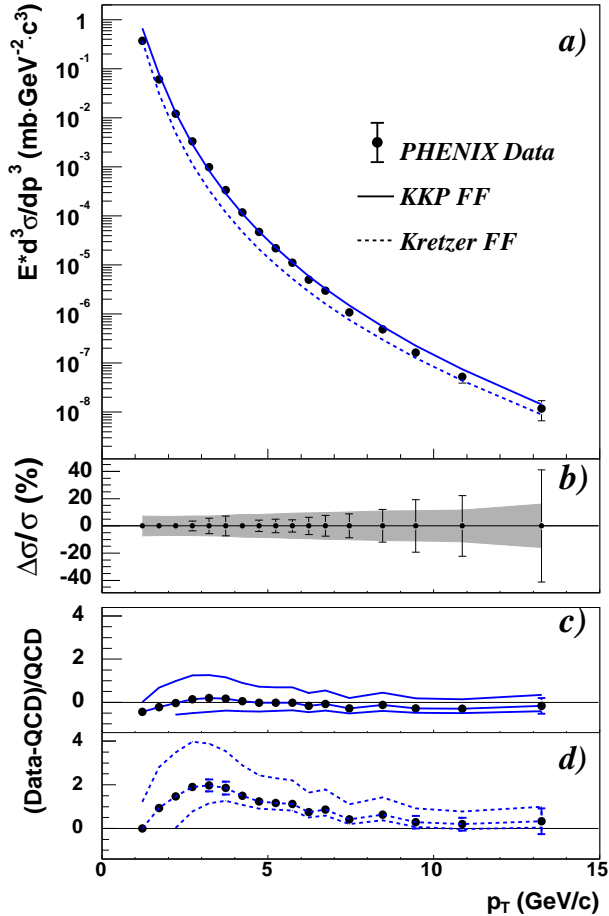}
 \caption{\it
Neutral pion production cross section as a function of $p_T$ measured in p+p collisions at $\sqrt{s_{NN}} = 200$~GeV
by the PHENIX experiment~\cite{PHENIX_pi0pp}.  Comparisons are shown to 
NLO pQCD calculations with two different fragmentation functions~\cite{KKP,Kretzer}.
    \label{fig:spectrum_pp} }
\end{figure}

In order to understand the complicated dynamics of a 
heavy ion collision, one
needs a calibrated, understood probe.  This can be provided by
hard-scattered partons, for which the fragmentation
dominates particle production at high $p_T$.  In particular, it is 
known that the fragmentation of partons 
into jets dominates the production of
hadrons above $p_T \sim 2$~GeV/c~\cite{Owens} in p+p collisions. 
For the
hard-scattered partons to serve as a probe of heavy ion collisions, 
a baseline for high $p_T$ particle production, 
at the same $\sqrt{s_{NN}}$, 
must first be established from p+p collisions.
Figure~\ref{fig:spectrum_pp} 
shows the neutral pion production cross section 
measured in p+p collisions at $\sqrt{s_{NN}} = 200$~GeV 
by the PHENIX experiment~\cite{PHENIX_pi0pp}.
The production cross section is well described by NLO pQCD 
calculations~\cite{KKP,Kretzer}.  
Therefore, the measurement in p+p collisions can provide
a baseline for measurements in Au+Au collisions.

In heavy ion collisions, the boundary between where hard and where soft 
production dominates the measured hadron yields is affected by
flow~\cite{teaney,flow} and possibly parton 
coalescence~\cite{hwa,mueller,greco,molnar}.  This complicates the 
high $p_T$ region between 2 and 4 GeV/c.  However,
in Run~II at RHIC ($\sqrt{s_{NN}} = 200$~GeV), hadrons have been 
measured up to $p_T \sim 10$~GeV/c in Au+Au 
collisions~\cite{PHENIX_pi0,PHENIX_h,STAR_h}.  
At such large transverse momenta, interactions are expected to be
incoherent even in
heavy ion collisions, and perturbative QCD calculations can reliably 
be used as a baseline for hadron production~\cite{vitev,ina,levai,xnwang}.
Thus, at sufficiently high transverse momenta,
we still have a calibrated probe for heavy ion collisions.
Hard scatterings provide a particularly valuable probe, in fact, 
because they occur early in the collision, leaving the hard-scattered 
partons sensitive to the properties of the collision medium. 
A hard scattering occurs on a timescale
of $\sim 1/p_T$, which for $p_T = 2$~GeV is approximately 0.1~fm/c.  This
is smaller than the time it takes for the system to 
equilibrate~\cite{Heinz_Kolb}. 
Therefore, the hard-scattered partons experience the entire 
space-time of the system and serve as a probe of the produced medium.  
It has been predicted that the hard-scattered partons
will reinteract in a deconfined medium of free color charges, 
losing much of their energy~\cite{gyu90}.
This would result in a suppression of the
high transverse momentum tail of the hadron spectrum, 
where the hadrons are likely to be the leading particles of jets, and
is known as ``jet quenching.''

\subsection{Centrality}

In a Au+Au collision, the properties of 
the produced medium depend on the centrality of the collision.
At large impact parameter, the overlap region between the 
two nuclei (or system size) is small; and 
at small impact parameter, the system size is large.  
Heavy ion collisions are generally binned into centrality
selections.  Such selections can be classified according to their geometry,
with a class of events having a mean number of participating nucleons
$N_{part}$ and a mean number of binary nucleon-nucleon collisions $N_{binary}$
determined by a Glauber~\cite{Glauber} model simulation. 
The maximum $N_{part}$ possible for a Au+Au collision is 394~(2x197), and
$N_{binary}$ exceeds 1000 for the most central events.
Physics observables are presented as a function of centrality, from the
most peripheral selection, 
corresponding to the events with the least amount of overlap
between the nuclei, to the most central selection,
corresponding to the events with the most overlap.
Centrality is expressed as a percentage 
of the total inelastic cross section.

\section{Scaling of Hadron Production}

At large transverse momenta, where the cross sections are small and
particles are expected to undergo incoherent interactions, 
a heavy ion (A+A) collision can be viewed simply as a 
superposition of binary nucleon-nucleon (N+N) collisions.
The scaling between systems of different sizes 
is the number of sources for hard parton scatterings.  Therefore,
to compare Au+Au collisions with p+p collisions, 
the scaling is the mean number
of nucleon-nucleon collisions $N_{binary}$, or alternately the Glauber 
nuclear overlap function $T_{AA}$.
We quantify the effects of the nuclear medium by the nuclear 
modification factor $R_{AA}$, which is the fraction of 
yields at high $p_T$ measured in A+A collisions to those measured in p+p 
collisions scaled by $N_{binary}$ (Eq.~\ref{eq:RAA_nbinary}) or $T_{AA}$
(Eq.~\ref{eq:RAA_taa}).  
\begin{eqnarray}
\nonumber R_{AA}(p_T) 
= \frac{(\rm Yield \; per \; A+A \; collision)}
{\langle N_{binary} \rangle ({\rm Yield \; per \; p+p \; collision})} \\
= \frac{ d^{2}N^{A+A}/dp_T d\eta }
{\langle N_{binary} \rangle  
( d^{2}\sigma^{p+p}/dp_T d\eta ) / \sigma^{p+p}_{inelastic} } \label{eq:RAA_nbinary} \\
= \frac{ d^{2}N^{A+A}/dp_T d\eta }
{\langle T_{AA} \rangle  
( d^{2}\sigma^{p+p}/dp_T d\eta ) }.
\label{eq:RAA_taa}
\end{eqnarray}
At low $p_T$, where soft production
dominates the measured yields, $R_{AA}$ is less than one because soft
production is expected to scale with the number of participants $N_{part}$,
rather than $N_{binary}$.
At transverse momenta sufficiently large to be in a hard-scattering 
regime and in the absence of any nuclear effects, $R_{AA}$ is expected 
to be equal to unity; in which case  
a A+A collision can be described as a 
superposition of binary N+N collisions.  A deviation from unity 
at high $p_T$
is a measure of the effect of the nuclear medium.  Previously measured 
nuclear effects include the ``Cronin'' effect~\cite{cronin}, an 
enhancement 
in the yields relative to binary scaling ($R_{AA} > 1$) attributed 
to $k_T$ broadening due to initial state multiple 
scattering~\cite{ptbroadening,pre_constantE}.
The predicted energy loss due to the dense medium, or jet quenching, is a 
suppression in the nuclear modification factor ($R_{AA} < 1$).

\section{Measurements of $R_{AA}$}

\begin{figure}[hbt]
\includegraphics[width=10cm]{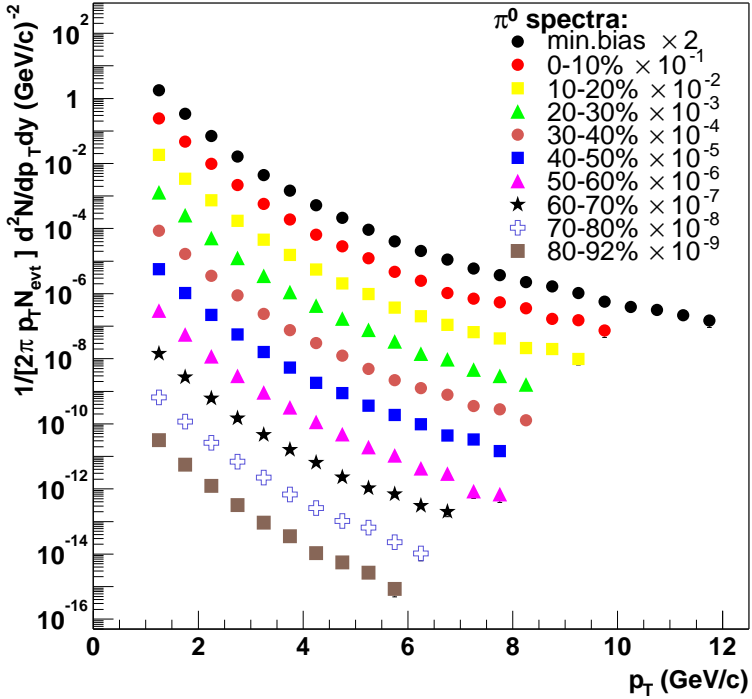}
 \caption{\it
Invariant neutral pion yields 
as a function of $p_T$ measured in Au+Au collisions 
at $\sqrt{s_{NN}} = 200$~GeV
by the PHENIX experiment~\cite{PHENIX_pi0}.  Spectra for different 
centrality selections are scaled by factors of 10 for presentation.
    \label{fig:spectra_AuAu} }
\end{figure}

The invariant yields of hadrons as a function of $p_T$
have been measured in Au+Au collisions at $\sqrt{s_{NN}} = $ 130, 200, 
and most recently 62~GeV.  
The charged hadron $p_T$ spectra have been 
measured by all 4 RHIC 
experiments~\cite{PHENIX_h,STAR_h,PHOBOS_h,dAu_BRAHMS}, and the neutral pion 
$p_T$ spectra
have been measured by the PHENIX experiment~\cite{PHENIX_supp,PHENIX_pi0}.
In Fig.~\ref{fig:spectra_AuAu}, 
the neutral pion $p_T$ spectra are shown for all centrality selections 
from the most central bin~(0-10\%) to the most 
peripheral (80-92\%), at $\sqrt{s_{NN}} = 200$~GeV.

\begin{figure}[hbt]
\includegraphics[width=10cm]{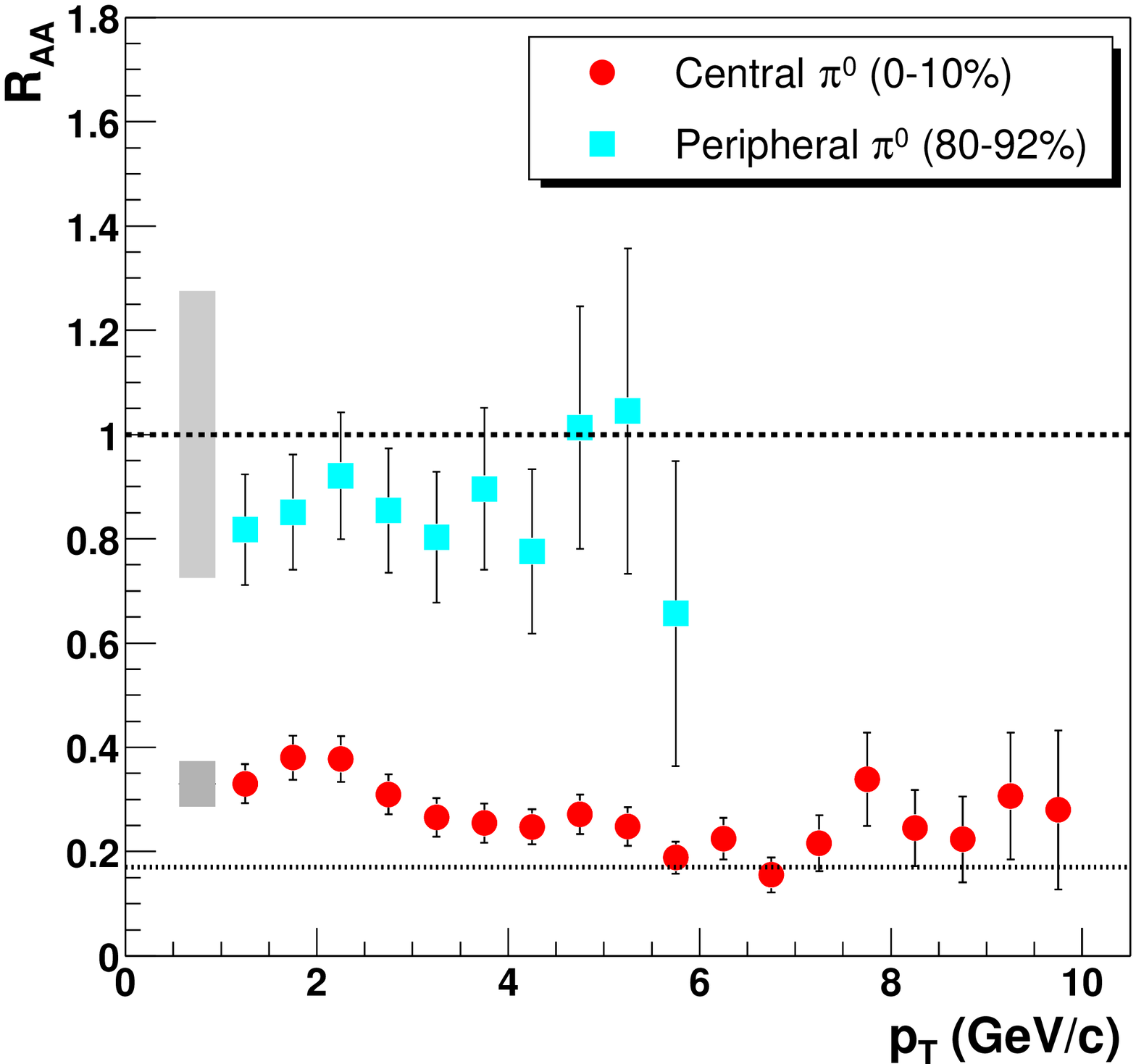}
 \caption{\it
The $R_{AA}$ for neutral pions 
as a function of $p_T$ measured in Au+Au collisions
by the PHENIX experiment~\cite{PHENIX_pi0}.  
The red points are for central events (0-10\% centrality), 
and the blue squares are for peripheral events (80-92\% centrality). 
The shaded bars indicate the fractional normalization 
error (shown as the fraction of the first data point), which is dominated
by the uncertainty in $\langle T_{AA} \rangle$.
    \label{fig:RAA} }
\end{figure}

\begin{figure}[hbt]
\includegraphics[width=10cm]{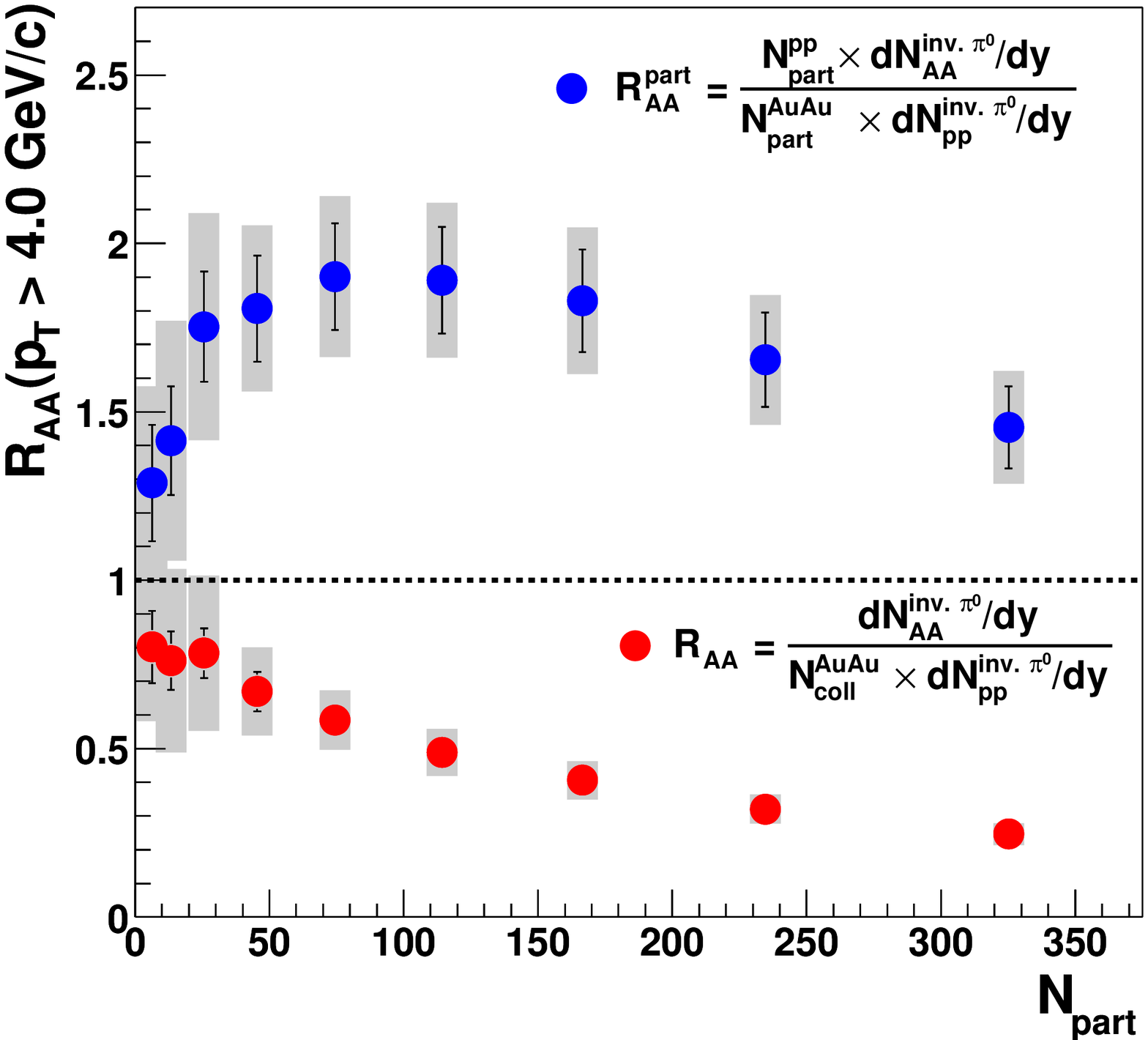}
 \caption{\it
The $R_{AA}$ for neutral pions 
as a function of $N_{part}$ (or centrality) measured in Au+Au collisions
by the PHENIX experiment~\cite{PHENIX_pi0} for integrated yields for
$p_{T} > 4$~GeV/c.  The lower red points show $R_{AA}$ as defined in Eq.~\ref{eq:RAA_nbinary}, while the upper blue points show $R_{AA}^{part}$, for which $N_{binary}$ in Eq.~\ref{eq:RAA_nbinary} is replaced by $N_{part}$.  The shaded bands around the points indicate the systematic errors on $N_{binary}$ or $N_{part}$.
    \label{fig:RAA_cent} }
\end{figure}

Figure~\ref{fig:RAA} shows the nuclear modification factor 
$R_{AA}$ (Eqs.~\ref{eq:RAA_nbinary} and~\ref{eq:RAA_taa}) 
for the most central and for the most peripheral Au+Au 
collisions at $\sqrt{s_{NN}}=200$~GeV.
The central data show a significant suppression (which was first observed at $\sqrt{s_{NN}} = 130$~GeV
at moderate $p_T$ of 2-4 GeV/c~\cite{PHENIX_supp}) 
of a factor of $\sim 4-5$ at high $p_T$; while
$R_{AA}$ for peripheral events is consistent with binary-scaling 
within errors (p+p yields scaled
by the number of $N_{binary}$ in the peripheral event sample).  
This indicates that the nuclear effects are large in central Au+Au 
collisions and small in peripheral collisions.
Shown in Fig.~\ref{fig:RAA_cent}, by the lower red points, 
is the evolution of $R_{AA}$
from the most peripheral to the 
most central collisions for yields integrated over $p_T > 4$~GeV/c, 
in terms of the
mean number of participating nucleons $N_{part}$ in a 
centrality selection.  
Here, one sees that the suppression gradually increases from
peripheral to central events.  

\begin{figure}[hbt]
\includegraphics[width=10cm]{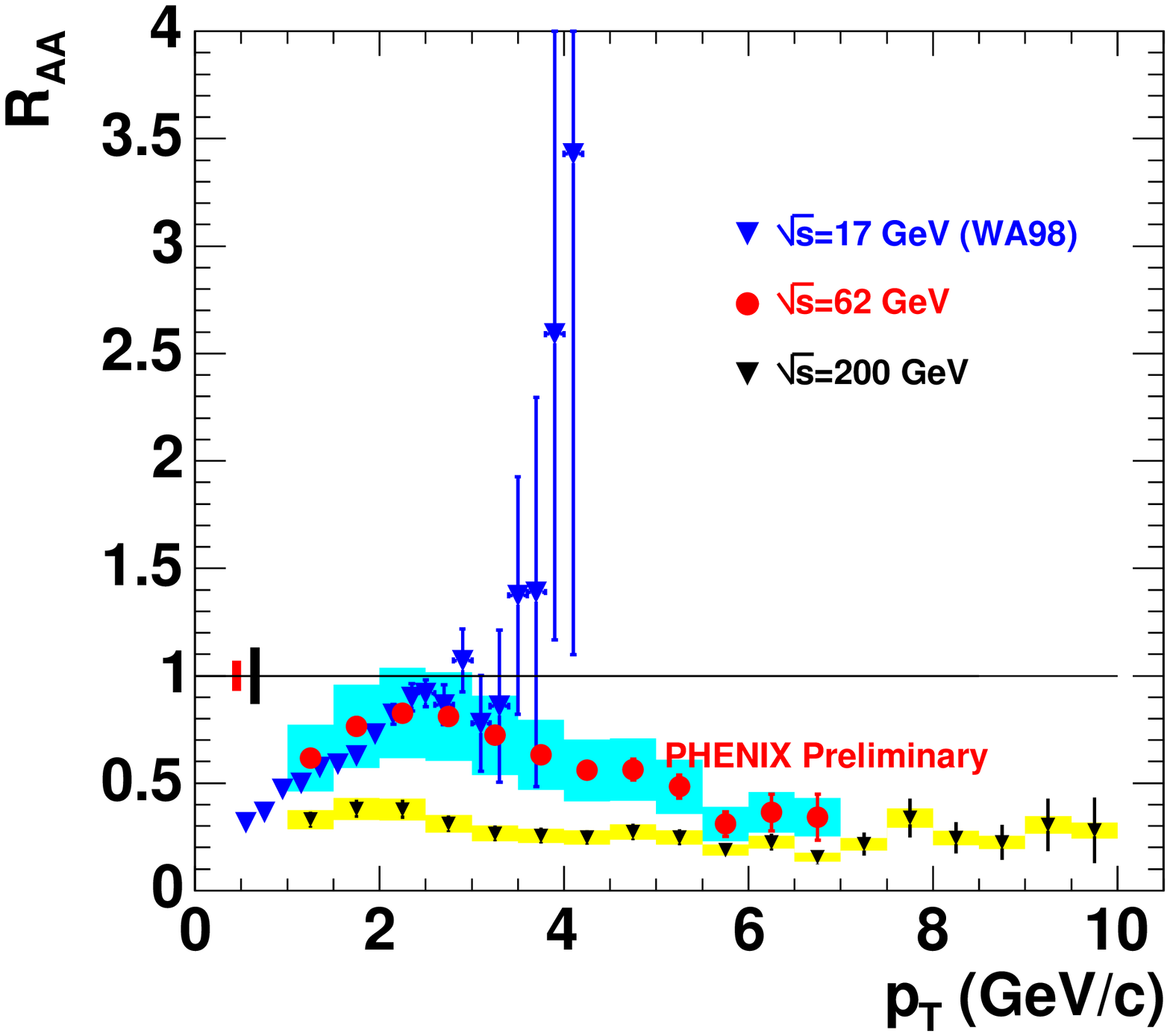}
 \caption{\it
$R_{AA}$ for neutral pions in central Au+Au collisions at RHIC at $\sqrt{s_{NN}} = 200$~GeV~\cite{PHENIX_pi0} and $\sqrt{s_{NN}} = 62$~GeV (PHENIX preliminary), and central Pb+Pb collisions at the SPS~\cite{wa98,David} ($\sqrt{s_{NN}} = 17$~GeV).  The boxes surrounding the data points indicate the
systematic uncertainties that are correlated in $p_T$. 
The black line at 1 shows the percent normalization 
error on the data for $\sqrt{s_{NN}} = 200$~GeV, and the red line for
$\sqrt{s_{NN}} = 62$~GeV.}
\label{fig:raa_ecomp}
\end{figure}

It is also useful to investigate the dependence of $R_{AA}$ on the
$\sqrt{s_{NN}}$.
Figure \ref{fig:raa_ecomp} shows the measured $R_{AA}$ for neutral pions 
in Pb+Pb collisions at $\sqrt{s_{NN}} = 17$~GeV at the SPS~\cite{wa98}
(re-analysis of p+p reference~\cite{David}) and
in Au+Au collisions at $\sqrt{s_{NN}} = 62$~GeV
and at $\sqrt{s_{NN}} = 200$~GeV~\cite{PHENIX_pi0} at RHIC. 
At the SPS, in Pb+Pb collisions, there is no apparent suppression, and 
any possible energy loss effect present in these collisions is
dominated by the Cronin effect.
At RHIC, for the first time, a suppression was observed in central Au+Au 
collisions consistent with the prediction for jet quenching. 
The suppression persisting up to $p_T = 10$~GeV/c, observed in Run II
for $\sqrt{s_{NN}} = 200$~GeV, was successfully predicted 
by theoretical calculations that invoke an energy loss proportional 
to the energy of the parton combined with shadowing and initial-state 
$k_T$-broadening~\cite{pre_nonconstantE1,pre_nonconstantE2}. 
A comparison with theoretical predictions is shown in 
Sec.~\ref{sec:theory}.  The most recent measurement at 
$\sqrt{s_{NN}} = 62$~GeV from Run IV is also included in the figure and 
shows a similar suppression for $p_T > 4-5$~GeV/c.

\section{Correlation Measurements to Detect Jets}

Due to the high multiplicity
environment, jets cannot be directly observed in a heavy ion collision.
When interpreting the modification of the particle spectra at high $p_T$
as an effect of the medium on the hard-scattered partons, one assumes
that the hadrons measured at high $p_T$ emanate from jets.
A method to detect the presence of jets in high multiplicity 
environment is 
via two-particle angular correlation measurements. 
To extract the jet signal from the correlation distributions,
other sources of correlation, such as flow and resonance decays, and the
combinatorial background need to be disentangled 
from the correlations due to jets.
The STAR experiment has made such a measurement~\cite{STAR_corr} and
found that the away-side jet disappears 
in central Au+Au collisions.  This is 
shown in the bottom right panel of Fig.~\ref{fig:dAu}.
The "near-side" correlations are those at $|\Delta \phi| \sim 0^{\circ}$, 
and the "away-side" are at $|\Delta \phi| \sim 180^{\circ}$.
Since the near-side contains the trigger particle,
the correlation function can be understood as a conditional probability.
When triggering on a high $p_T$ particle, the near-side correlation
looks very similar to the correlation due to jets measured in 
p+p collisions 
(black histogram); but there is no jet correlation seen on the away-side
in central Au+Au events,
contrary to what is measured in p+p collisions.  A possible interpretation
is that the hadrons that we measure at high $p_T$ 
in central Au+Au collisions come
from hard scatterings near the surface of the system.  This would allow one
jet to escape the medium; while the corresponding jet must traverse the
dense medium where it could interact, lose much its energy, 
and become lost in the low $p_T$ soft part of the spectrum. 

Although both observations that were made in central
Au+Au collisions at RHIC (the suppression in the hadron 
yields at high $p_T$ and the disappearance of the away-side jet) seem to 
indicate that the hard-scattered partons lose energy in the dense medium,
the effect of the ``cold'' nuclear medium (p+A or d+A collisions) 
is essential to rule out initial-state nuclear effects.

\section{Results from d+Au Collisions}

\begin{figure}[htb]
\includegraphics[width=13cm]{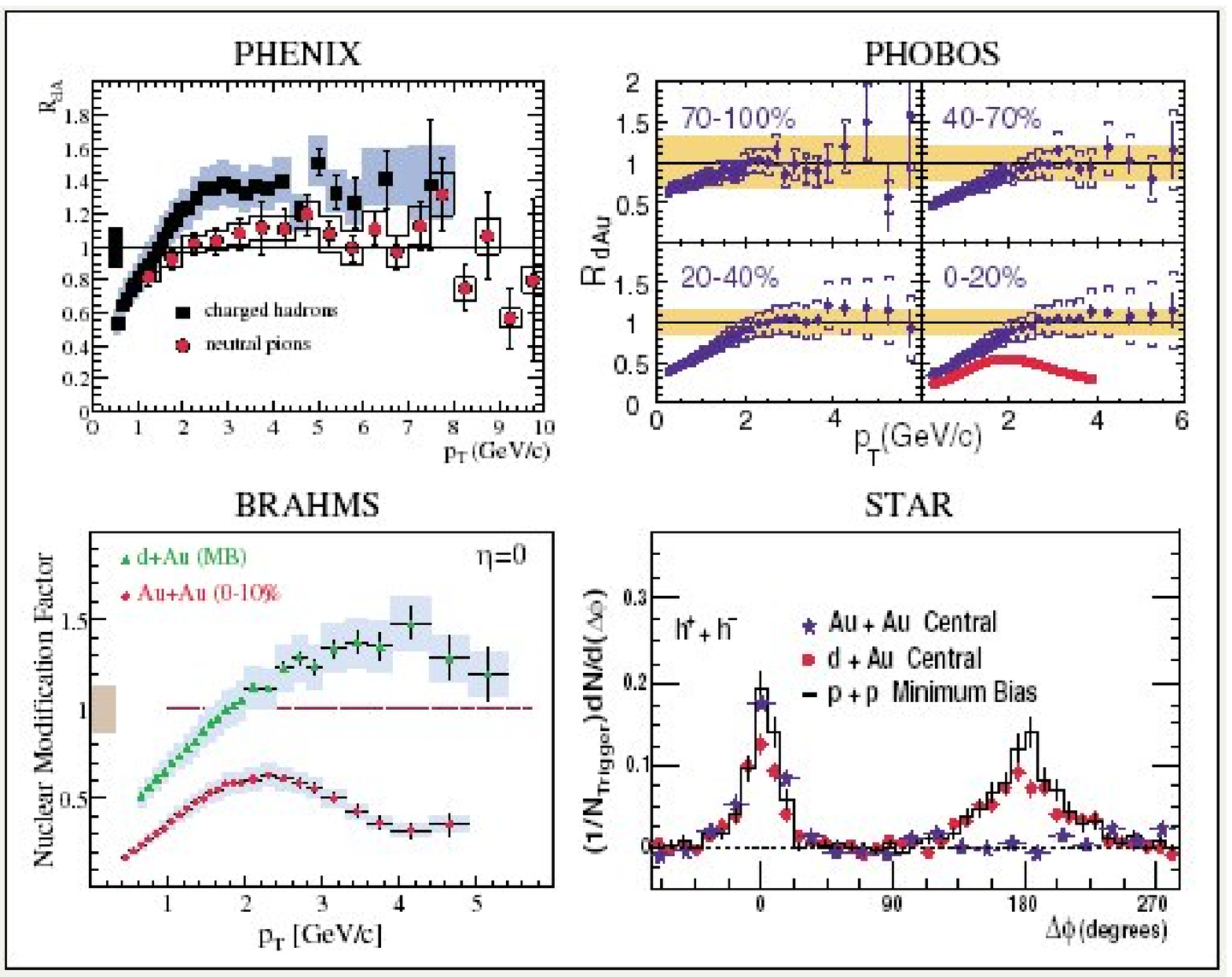}
 \caption{\it
Results of measurements in d+Au collisions 
from all 4 RHIC experiments as a comparison to the novel effects
observed in central Au+Au collisions.  The two upper and the lower left panel
show $R_{AA}$ vs. $p_T$, and the bottom right panel shows the angular 
correlation measurement due to jets.  The upper left shows
$R_{AA}$ for neutral pions and charged hadrons measured in 
d+Au collisions by PHENIX \cite{dAu_PHENIX}.  
The upper right shows $R_{AA}$ for charged hadrons measured 
in different centrality selections in d+Au collisions by PHOBOS 
\cite{dAu_PHOBOS}, where the most central in d+Au collisions
is compared to the measurement in central Au+Au collisions (the red line).
The lower left is $R_{AA}$ in d+Au collisions compared to central 
Au+Au collisions measured by BRAHMS \cite{dAu_BRAHMS}.  
The lower right is the jet correlation
signal measured in p+p collisions, d+Au collisions, and central Au+Au 
collisions by STAR \cite{dAu_STAR}.}
\label{fig:dAu} 
\end{figure}

In Run~III at RHIC, high $p_T$ particle production was studied in 
d+Au collisions.  To distinguish initial-state effects from final-state 
effects on the suppression observed in central Au+Au collisions, 
d+Au collisions provide a good comparison 
experiment with only initial-state nuclear 
effects.  Figure~\ref{fig:dAu} shows the results from all 4 RHIC 
experiments \cite{dAu_PHENIX,dAu_PHOBOS,dAu_BRAHMS,dAu_STAR}.  
From these results, the consistent conclusion is that the 
suppression observed in central Au+Au collisions is not observed in d+Au 
collisions.  Similarly, the disappearance of the away-side jet 
(bottom right panel) is also not observed in d+Au collisions.
This indicates
that the novel phenomena observed in central Au+Au collisions at RHIC are 
due to effects of the medium produced in these collisions, rather than 
initial-state nuclear effects.

\section{Comparison to Theory \label{sec:theory}}

\begin{figure}[htb]
\includegraphics[width=13cm]{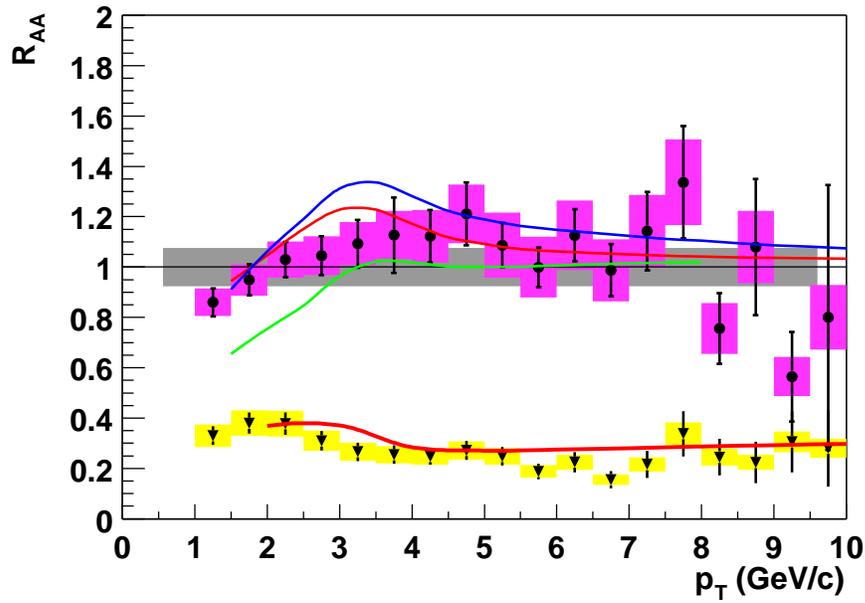}
 \caption{\it
The measured $R_{AA}$ as a function of $p_T$ compared with theoretical
predictions for central Au+Au collisions~\cite{theory1} and d+Au collisions~\cite{theory2}.  The boxes around the data points indicate the systematic
errors correlated in $p_T$, and the shaded grey band at 1 indicates the percent normalization error.
    \label{fig:RAA_theory} }
\end{figure}

The measurements of high $p_T$ hadron production in Au+Au and 
d+Au collisions have been compared with theoretical calculations.
Figure~\ref{fig:RAA_theory} shows the
model prediction for $R_{AA}$ in central Au+Au events (lower red line), 
which includes the effect of energy loss of hard-scattered partons.  
It is in good agreement with the data.  For the comparison to 
the measurement in d+Au collsions,
the predictions from the same model include only initial-state effects, 
shown by the different lines (for varying initial-state effects) 
near and above $R_{AA} = 1$.  The model results for d+Au 
collisions are also in reasonable agreement with the data.
This supports the conclusion 
that a final-state nuclear medium effect, such as parton energy loss, 
is necessary to describe the suppression observed in central 
Au+Au collisions.

\clearpage

\section{Conclusions}

At RHIC, one of
the most significant observations thus far is the depletion of
high $p_T$ hadrons in central Au+Au collsions.  Such a suppression
is consistent with predictions in which hard-scattered partons lose energy
in a dense medium.  Complementing this observation is the discovery of the
disappearance of the 
away-side jet in central Au+Au collisions, which 
can also be explained by
interactions of the hard-scattered partons in a dense medium. 
As a reference for the effect of cold nuclear matter on hadron spectra at
high $p_T$ and on correlations due to jets, 
the same measurements were made in d+Au
collisions.  The charged hadron and neutral pion spectra were found not 
to be suppressed relative to the binary-scaled spectra measured in 
p+p collisions, and the away-side jet was found not to disappear.  

The measurements of $R_{AA}$ are in agreement with theoretical predictions
both for central Au+Au and for d+Au collisions.  The model includes
the effect of energy loss in a dense medium, as well as initial-state nuclear effects, in central Au+Au collisions, 
and only initial-state effects in d+Au collisions.
 
Both the results from d+Au collisions and the comparisons to theory support 
the conclusion that the suppression of high $p_T$ hadrons is an effect due
to the produced medium in central Au+Au collisions.


\begin{thebibliography}{99}
\bibitem{lattice}
See E. Laermann and O. Philipsen,
Ann. Rev. Nuc. Part. Sci. {\bf 53}, 163 (2003), for a recent review.
\bibitem{Owens} 
J.F.~Owens {\it et al.}, Phys. Rev. D{\bf 18}, 1501 (1978).
\bibitem{PHENIX_pi0pp}
S.S.~Adler {\it et al.}, Phys. Rev. Lett. {\bf 91}, 241803 (2003).
\bibitem{KKP}
B.A.~Kniehl {\it et al.}, Nucl. Phys. B{\bf 597}, 337 (2001).
\bibitem{Kretzer}
S.~Kretzer {\it et al.}, Phys. Rev. D{\bf 62}, 054001 (2000).
\bibitem{teaney}
D.~Teaney, J.~Lauret and E.V.~Shuryak, nucl-th/0110037.
\bibitem{flow}
See P.F.~Kolb and U.~Heinz, nucl-th/0305084, for a recent review.
\bibitem{hwa}
R.~C.~Hwa and C.~B.~Yang, Phys. Rev. C{\bf 67}, 034902 (2003).
\bibitem{mueller}
R.~J.~Fries, B.~M\"uller, C.~Nonaka and S.~A.~Bass, Phys. Rev. Lett.{\bf 90}, 202303 (2003).
\bibitem{greco}
V.~Greco, C.~M.~Ko and P.~Levai, Phys. Rev. Lett.{\bf 90}, 202302 (2003).
\bibitem{molnar}
D.~Molnar and S.A.~Voloshin, Phys. Rev. Lett.{\bf 91}, 092301 (2003).
\bibitem{PHENIX_pi0}
S.S.~Adler {\it et al.}, Phys. Rev. Lett.{\bf 91}, 072301 (2003). 
\bibitem{PHENIX_h}
S.S.~Adler {\it et al.}, Phys. Rev. C{\bf 69}, 034910 (2004).
\bibitem{STAR_h}
J.~Adams {\it et al.}, Phys. Rev. Lett.{\bf 91}, 172302 (2003). 
\bibitem{vitev} 
I.~Vitev and M.~Gyulassy, Phys. Rev. Lett{\bf 89}, 252301 (2002).
\bibitem{ina} 
S.~Jeon, J.~Jalilian-Marian and I.~Sarcevic, Phys. Lett. B{\bf 562}, 45 (2003).
\bibitem{levai} 
G.G.~Barnafoldi {\it et al.}, nucl-th/0212111.
\bibitem{xnwang} 
X.~N.~Wang, Nucl. Phys. A{\bf 715}, 775 (2003).
\bibitem{Heinz_Kolb}
U.W.~Heinz and P.F.~Kolb, Nucl. Phys. A{\bf 702}, 269-280 (2002).
\bibitem{gyu90} 
M.~Gyulassy and M.~Pl\"umer, Phys. Lett. B{\bf 243}, 432 (1990);
X.N.~Wang and M.~Gyulassy, Phys. Rev. Lett.{\bf 68}, 1480 (1992);
R.~Baier {\it et al.}, Phys. Lett. B{\bf 345}, 277 (1995).
\bibitem{Glauber} 
R.J.~Glauber and G.~Matthiae, Nucl. Phys. B{\bf 21}, 135 (1970).
\bibitem{cronin} 
D.~Antreasyan {\it et al.}, Phys. Rev. D{\bf 19}, (1979) 764.
\bibitem{ptbroadening} 
M.~Lev and B.~Petersson, Z. Phys. C{\bf 21}, (1983) 155; 
T.~Ochiai {\it et al.}, Prog. Theor. Phys. {\bf 75}, (1986) 288.
\bibitem{pre_constantE} X.N.~Wang, Phys. Rev. C{\bf 61}, 064910 (2000).
\bibitem{PHOBOS_h}
B.~Back {\it et al.}, Phys.Lett.B{\bf 578}, 297-303 (2004). 
\bibitem{dAu_BRAHMS}
I.~Arsene {\it et al.}, Phys. Rev. Lett.{\bf 91}, 072305 (2003).
\bibitem{PHENIX_supp} 
K.~Adcox {\it et al.}, Phys. Rev. Lett.{\bf 88}, 022301 (2002).
\bibitem{wa98} 
M.M.~Aggarwal {\it et al.}, Eur. Phys. J{\bf 23}, 225 (2002).
\bibitem{David}
D.~d'Enterria, nucl-ex/0403055 (re-analysis of p+p reference at $\sqrt{s_{NN}} = 17$~GeV).
\bibitem{pre_nonconstantE1} 
P.~Levai {\it et al.}, Nucl. Phys. A{\bf 698}, 631 (2002).
\bibitem{pre_nonconstantE2} 
I.~Vitev and M.~Gyulassy, Phys. Rev. Lett.{\bf 89}, 252301 (2002).
\bibitem{STAR_corr}
C.~Adler {\it et al.}, Phys. Rev. Lett.{\bf 90}, 082302 (2003).
\bibitem{dAu_PHENIX}
S.S.~Adler {\it et al.}, Phys. Rev. Lett.{\bf 91}, 072303 (2003).
\bibitem{dAu_PHOBOS}
B.B~Back {\it et al.}, Phys. Rev. Lett.{\bf 91}, 072302 (2003).
\bibitem{dAu_STAR}
J.~Adams {\it et al.}, Phys. Rev. Lett.{\bf 91}, 072304 (2003).
\bibitem{theory1}
I. Vitev and M. Gyulassy, Nucl. Phys. A{\bf 715}, 779-782 (2003).
\bibitem{theory2}
I. Vitev, Phys. Lett B{\bf 562}, 36-44 (2003).
\end{thebibliography}
\end{document}